\documentstyle[preprint,aps]{revtex}

\begin{document}

\draft

\title{Simple model for the phase coexistence and electrical conductivity
of alkali fluids }

\author{
P. Tarazona$^1$, E. Chac\'on$^2$, and J. P. Hernandez$^3$ }

\address{
$^1$Departamento de F\'isica de la Materia Condensada (C-XII),
Universidad Aut\'onoma de Madrid, E-28049 Madrid, Spain }

\address{
$^2$Instituto de Ciencia de Materiales, Consejo Superior de
Investigaciones Cient\'ificas and \\  Departamento de
F\'isica Fundamental, Universidad Nacional de Educaci\'on a Distancia,
Apartado 60141, E-28028 Madrid, Spain}

\address{
$^3$Department of Physics and Astronomy, University of North Carolina,
Chapel Hill NC 27599-3255 USA }

\date{\today}
\maketitle

\begin{abstract}
We report the first theoretical model for the alkali fluids which yields a
liquid-vapor phase coexistence with the experimentally observed features
and electrical conductivity estimates which are also in accord with
observations. We have carried out a  Monte Carlo simulation for a lattice
gas model which allows an integrated study of the structural,
thermodynamic, and electronic properties of metal-atom fluids. Although
such a technique is applicable to both metallic and nonmetallic fluids,
non-additive interactions due to valence electron delocalization are a crucial
feature of the present model.  \end{abstract}


        Until recently little has been known regarding metal-atom fluids
near the high temperatures and pressures characteristic of their
liquid-vapor critical points. However, such knowledge is not only of
considerable scientific interest but is also required for potential
technological applications near those conditions. The electronic and
thermodynamic properties of such fluids are intimately related and
knowledge of these properties is essential in determining the right
material for a given application. Such considerations have motivated
experimental studies of those materials with the lowest critical temperatures:
Hg (1751 K), Cs (1924 K), and Rb (2017 K) \cite {Gener1}, with data on K
(2178 K) now being available \cite {Hohl}. The data have become precise and
reliable in the last decade and span thermodynamic and electrical
measurements under the same conditions. Such data show that the
liquid-vapor coexistence curve of metal-atom fluids are different
from those of Lennard-Jones-like ones \cite {Hohl,Jungst}. For example,
the liquid and vapor branches of the coexistence curves are strongly
asymmetric and the rectilinear diameter law breaks down over a large
temperature range, not only very close to the critical points. These
materials also undergo a metal-nonmetal (M-nM) transition. This body of data,
however, still seeks microscopic theoretical foundations \cite {Stratt}.
This paper presents a simple model which appears to contain the basic
ideas required to reproduce, in a unified manner, the peculiar
characteristics observed in the alkali fluids.

        The goal of the present work - understanding the structural,
thermodynamic, and electronic properties of metal-atom fluids - poses a
considerable scientific challenge. Its various aspects are coupled since
it is the electrons which determine interatomic interactions and thus the
structure and thermodynamic data. The ionic structure, in turn, determines
the electronic properties. In the study of metal-atom fluids it is
difficult to impose a structure, since it is so intimately related to
electronic effects and there is no long-range symmetry to simplify the
problem. Also, because in such materials the interactions are not pairwise
ones for intermediate densities, due to electron delocalization over some
regions, the problem is more complicated than that for fluids in which
the interactions are not state dependent. Similarly, as these materials
also undergo a M-nM transition, the traditional techniques
used to study free-electron-like fluids are not applicable over many of the
conditions of interest. The microscopic theory required for these
materials should seek to explain the essential interdependence of
thermodynamic, structural, and electronic properties. Previous theoretical
efforts to comprehensively explain the available experimental data on
metal-atom fluids have been sparse. The points of view taken were usually
based on the limiting cases of either a metallic dense liquid or solid, or
alternatively of a nonmetallic dilute vapor. Attempts were then made to
describe the fluid, or some of its properties, in a limited density and
temperature range \cite {Hafner,Minchin,Hernan1}. General arguments based
on electron correlation effects (Hubbard model) and/or disorder induced
localization (Anderson model) \cite {Lee} are useful to study the M-nM
transition in systems with frozen ionic structure but probably not in
metal-atom fluids.

A first step towards a comprehensive theoretical treatment attempted the
extension of concepts and techniques of plasma physics to treat both liquid
and vapor phases, and then tried to introduce the required neutral atoms
and small clusters \cite {Gener2}. We also proposed such an approach
\cite {Chacon,Hernan2} and showed that a toy model gives a liquid-vapor
coexistence and a critical point with some correct features, but we
were far from reproducing the peculiar coexistence curve of alkali fluids.
In recent work \cite {Rostoc}, we followed this approach including a
quantitatively good description of the plasma: a standard description of liquid
metals near their triple point \cite {Shimoji}. Extension of this
treatment to high temperature and low density gave a plasma liquid-vapor
coexistence with a very high critical temperature (around four times the
experimental value) and a very different shape for the coexistence curve
than that observed. We then extended the model to allow for chemical
coexistence of neutral atoms with the plasma. Virial ion-atom
interactions were used, instead of their neglect as in the previous toy model.
However, reasonable values of the parameters did not improve the previous
plasma results. We have concluded that a mean-field theory is not capable of
reproducing the structure or phase diagram of the alkali fluids \cite{Rostoc}.

An unanswered question, in the above approach, is: why do some fraction of
the valence electrons choose to be bound in atoms while others are
delocalized, at fixed temperature and density? The answer must lie in a
hitherto ignored underlying structure. An important clue is that clustering
effects are strongly enhanced for metal-atom systems, compared to
nonmetallic ones, due to their high cohesion which arises from the valence
electron delocalization over the cluster. In contrast, to retain its valence
electron an atom should have no near neighbors to which that electron can
be favorably delocalized. We present here a simple lattice gas model which
includes such considerations and yields results showing the observed
peculiarities of the alkali fluids.

In our model, we allow the ions to partially occupy the sites of a body
centered cubic (bcc) lattice, which has the same maximum number of nearest
neighbors as do real metals. The lattice parameter is determined by the
condition that the maximum density $ \rho _{0} $  be equal to that of the
liquid metal at its triple point. The energy, in a mean-field treatment of the
model, is obtained from the energy per ion $u( \rho)$ in a normal treatment
for a liquid metal with mean density $ \rho $ \cite {Shimoji}. That is,
using a jellium reference system for the delocalized valence electrons
yields kinetic, exchange, and correlation energy contributions, and a
linear response function. Also, a hard-sphere reference for the ions gives
a pair distribution function. Finally, a pseudopotential is associated with
the ions (Ashcroft, empty core, fitted to the liquid metal conductivity at
the melting point) and the screened ion-electron and ion-ion interactions
(minus reference system effects) are treated by perturbation theory. This
is a Gibbs-Bogoliubov variational approximation. The electronic effects
are treated at zero temperature since no appreciable changes result from
including the temperature effects in the kinetic and exchange energies. The
results with the mean-field lattice model (Table I) are very close to those
of the previous, continuous, plasma model \cite {Rostoc}.

The crucial point, however, is to calculate the energy of the system taking
into account the strong inhomogeneities due to clustering. We now propose to
get the energy per ion using the mean-field energy  $ u( \hat { \rho  } ) $
which is to be evaluated at a local average density $ \hat {\rho}  $. This
density takes into account the ion and its environment, through the
occupation of its nearest-neighbor lattice sites. Thus, $ \hat {\rho} =
(n+1) \rho _{0} /9 $ and n is the number of occupied nearest-neighbor
sites; it takes values from 0 to 8. For ions without any nearest neighbors
($ n=0 $), this energy, calculated as in a plasma, is compared with that of
a free atom ( i.e. minus the ionization energy ) and in a variational
spirit the lower value is chosen. Although the energy difference is small,
the atomic state is lower for all the alkalis. This approach thus includes
the structurally-based possibility of valence electron localization, to
form atoms, or some degree of delocalization. For cesium we take $ \rho _{0}
= 1.84 \ g/ cm^3 $ and obtain the following energies per ion (in eV):

\vspace{0.2cm}
\noindent
\begin{tabular}{cccccccccc}
   $ n $ & 0 & 1 & 2 & 3& 4 & 5 & 6 & 7 & 8  \\
  $ u  $ &
 -3.89 & -4.20 & -4.42 & - 4.58 & -4.71 & -4.81 & -4.90 & -4.99 &
-5.06 .\\
\end{tabular}
\vspace{0.2cm}

\noindent
This simple model for the configuration energy contains a basic difference
with the pairwise interactions of usual lattice models: non-additive
interactions with neighboring ions due to valence electron delocalization.
Also, such interactions are misrepresented in a mean-field treatment of a
metal-atom fluid if the density fluctuations become important.

We present the results of Monte Carlo simulations for this model, applied to
cesium and potassium. We have carried out simulations within a cube with
twelve bcc cells on each side (3456 sites). Some results have been checked
using a larger cube (8192 sites), without appreciable changes in the
results. The simulations, carried out at fixed temperature and chemical
potential, give the equilibrium density and internal energy of the system.
The pressure is obtained by thermodynamic integration. Structural features
such as a nearly linear decrease in the average coordination of ions
in the expanded coexisting liquid result from this calculation; that
feature agrees with what has been deduced from analysis of neutron scattering
data \cite {Winter}. In Fig. 1, we show the coexistence curves, in critical
reduced units, obtained for cesium along with the experimental result.
Using the present method, the shape of the coexistence curve is in good
agreement with experiments, recovering the strong liquid-vapor asymmetry.
Although the size of our simulation box prevents a closer approach to the
critical point than about one per cent, our results clearly show a
non-classical critical behavior, in agreement with experiment and previous
theoretical predictions \cite {Gold1,Gold2}.  Near the critical point, the
diameter $ \rho _{d} = ( \rho _{L} + \rho _{V} ) / 2 \rho _{C} $ is
indicated in the figure and clearly shows the nonlinear behavior: $ \rho
_{d} -1 \propto | (T_{C} - T)/T_{C} | ^{ 1- \alpha } $. Our estimate for $
1 - \alpha$, $  0.8 \pm 0.1 $, is compatible with  the experimental value
\cite {Jungst} of $ 0.87 \pm 0.03$. We have also calculated the phase
diagram for potassium; in reduced critical units it is very similar to that
of cesium. In agreement with experiments, our results for different alkali
fluids give similar reduced phase diagrams.

The critical temperatures in our model are higher than those of
the real fluids, as one would expect from a lattice-gas treatment of the
configurational entropy. However, the results in table I show that the
correlation effects included in the present model produce a very large
decrease of the critical temperature compared with the previously mentioned
mean-field approach. Also, the critical densities and pressures are brought
into reasonable agreement with experiments. Moreover, the relative critical
parameters of cesium and potassium in our model are very similar to the
relative experimental ones.

We have next explored electronic properties related to structure. The
experimental signature of a M-nM transition in these systems is a
decrease of several orders of magnitude in the conductivity of the expanded
fluid metal. In our model, such behavior is driven by the percolation of
the ionic cluster structure, rather than by considerations such as those of
Mott or Anderson. From typical configurations of our Monte Carlo simulation
for cesium, we have obtained the cluster structure at different temperatures
and pressures. The conductivity was estimated following the Kirchoff's law
model proposed by Nield et al. \cite {Nield}, with arbitrary units
fitting the experimental conductivity at the maximum density. The results
in Fig.2 show the conductivity estimated in this manner along the critical
isotherm versus the pressure, in excellent agreement with experiment \cite
{Gener1}. The figure also shows the calculated conductivity versus density
at the critical temperature, and that arising from a random filling of
the lattice. The difference clearly shows the effects of clustering. The
percolation density (at which the conductivity goes to zero) at $ T _{c} $
is less than half that obtained with random occupation.

We have presented a model allowing a unified study of the structural,
thermodynamic, and electronic properties of metal-atom fluids. The model
takes into account non-additive interactions, due to valence electron
delocalization, and it is studied with a Monte Carlo simulation which goes
beyond mean-field, as is required. Although the model is a very simplified
representation of a metal-atom fluid, comparisons with a range of
experimental results show that it contains the basic ingredients to allow
reproduction of the peculiar behavior observed in these systems. These
peculiarities include the metal-nonmetal and liquid-vapor transitions
and the connection between ionic and electronic structures. Our
results for the critical temperatures are still too high, though the
densities and pressures are acceptable. It is reasonable to expect that,
with a more realistic description of the fluid entropy, a similar model
(though more cumbersome to study) would also give a quantitatively good
result for the critical temperatures. Our simplified model appears to allow
a unified understanding of the peculiar characteristics of the alkali fluids.

This work was supported by the Direcci\'on General de Investigaci\'on
Cient\'ifica y T\'ecnica (Spain) under Grant PB91-0090 and the Instituto
Nicolas Cabrera.

\newpage
\begin{figure}
\caption{
Liquid-vapor coexistence curve of cesium. Filled circles: present MC
simulation; full line: fit to the experimental results of ref. [3].
The MC simulation diameter function, $\rho_{d} $,
is also plotted (crosses). The triangles show our metal-nonmetal transition
( i.e. the percolation line), the dotted line is a guide to the eye.}
\label{fig1}
\end{figure}

\begin{figure}
\caption{
On the left we show the electrical conductivity $ \sigma $
versus the reduced pressure along the critical isotherm:
crosses - our model
; open circles - experimental values of ref. [1],
both normalized to the conductivity
$ \sigma _{0} $ at our highest density $ \rho _{0} $.
On the right we
show our calculated electrical conductivity but now
versus the fractional occupation of the lattice;
crosses are our results for $ T = T _{c} $ and triangles for the random
filling of the lattice. }
\label{fig2}
\end{figure}

\newpage
\begin{table}[h]
\noindent
\label{table1}
\caption{
Comparison of the calculated critical conditions, using lattice gas
mean-field and the present theory, with the experimental results of
ref. [2] and [3]. The temperature $ T $ is in Kelvin, the pressure $ P $ in
bar, and the density $ \rho $ in $ g $ $ cm^{-3} $.}

\begin{tabular}{|c|ccc|ccc|} \hline
& \multicolumn{3}{c|}{ CESIUM } &
\multicolumn{3}{c|}{ POTASSIUM } \\
  & $ T_ {c} $ & $ \rho _{c}$ &  $ P _{c} $ &
 $ T_ {c} $ & $ \rho _{c}$ &  $ P _{c} $  \\
\hline
 & & & & & &  \\
 Mean Field  & 6730 & 0.03 & 9.6 & 7890 & 0.018 & 19    \\
 Our Theory & 3600 & 0.50 &  100 & 3920 & 0.21 & 170    \\
 Experiment & 1924 & 0.38 & 92.5 & 2178 &0.18& 148   \\
 & & & & & &  \\
\hline
\end{tabular}
\end{table}

\end{document}